\documentclass[twocolumn]{aastex63}

\received{February 12, 2021}
\revised{March 3, 2021}
\accepted{March 4, 2021}

\submitjournal{ApJL}

\shorttitle{OH emission in the day-side spectrum of WASP-33b}
\shortauthors{S.K. Nugroho et al.}

\graphicspath{{./}{}}

\begin{document}

\title{First Detection of Hydroxyl Radical Emission from an Exoplanet Atmosphere:\\High-dispersion Characterization of WASP-33b using Subaru/IRD \footnote{Based on data collected at Subaru Telescope, which is operated by the National Astronomical Observatory of Japan.}}

\correspondingauthor{Stevanus K. Nugroho}
\email{stevanus.nugroho@nao.ac.jp, skristiantonugroho@gmail.com}

\author[0000-0003-4698-6285]{Stevanus K. Nugroho}
\affiliation{Astrobiology Center, NINS, 2-21-1 Osawa, Mitaka, Tokyo 181-8588, Japan}
\affiliation{National Astronomical Observatory of Japan, NINS, 2-21-1 Osawa, Mitaka, Tokyo 181-8588, Japan}
\affiliation{School of Mathematics and Physics, Queen's University Belfast, University Road, Belfast, BT7 1NN, United Kingdom}

\author[0000-0003-3309-9134]{Hajime Kawahara}
\affiliation{Department of Earth and Planetary Science, The University of Tokyo, Tokyo 113-0033, Japan}
\affiliation{Research Center for the Early Universe, School of Science, The University of Tokyo, Tokyo 113-0033, Japan}

\author[0000-0002-9308-2353]{Neale P. Gibson}
\affiliation{School of Physics, Trinity College Dublin, The University of Dublin, Dublin 2, Ireland}

\author[0000-0001-6391-9266]{Ernst J. W. de Mooij}
\affiliation{School of Mathematics and Physics, Queen's University Belfast, University Road, Belfast, BT7 1NN, United Kingdom}

\author[0000-0003-3618-7535]{Teruyuki Hirano}
\affiliation{Astrobiology Center, NINS, 2-21-1 Osawa, Mitaka, Tokyo 181-8588, Japan}
\affiliation{National Astronomical Observatory of Japan, NINS, 2-21-1 Osawa, Mitaka, Tokyo 181-8588, Japan}

\author[0000-0001-6181-3142]{Takayuki Kotani}
\affiliation{Astrobiology Center, NINS, 2-21-1 Osawa, Mitaka, Tokyo 181-8588, Japan}
\affiliation{National Astronomical Observatory of Japan, NINS, 2-21-1 Osawa, Mitaka, Tokyo 181-8588, Japan}
\affiliation{Department of Astronomy, School of Science, The Graduate University for Advanced Studies (SOKENDAI), 2-21-1 Osawa, Mitaka, Tokyo 181-8588, Japan}

\author[0000-0003-3800-7518]{Yui Kawashima}
\affiliation{Cluster for Pioneering Research, RIKEN, 2-1 Hirosawa, Wako, Saitama 351-0198, Japan}

\author[0000-0003-1298-9699]{Kento Masuda}
\affiliation{Department of Earth and Space Science, Osaka University, Osaka 560-0043, Japan}

\author[0000-0002-7704-0153]{Matteo Brogi}
\affiliation{Department of Physics, University of Warwick, Coventry CV4 7AL, UK}
\affiliation{INAF—Osservatorio Astrofisico di Torino, Via Osservatorio 20, I-10025, Pino Torinese, Italy}
\affiliation{Centre for Exoplanets and Habitability, University of Warwick, Gibbet Hill Road, Coventry CV4 7AL, UK}

\author[0000-0002-4125-0140]{Jayne L. Birkby}
\affiliation{Astrophysics, Department of Physics, University of Oxford, Keble Road, Oxford OX1 3RH, UK}

\author[0000-0002-9718-3266]{Chris A. Watson}
\affiliation{School of Mathematics and Physics, Queen's University Belfast, University Road, Belfast, BT7 1NN, United Kingdom}

\author[0000-0002-6510-0681]{Motohide Tamura}
\affiliation{Department of Astronomy, Graduate School of Science, The University of Tokyo, 7-3-1 Hongo, Bunkyo-ku, Tokyo 113-0033, Japan}
\affiliation{Astrobiology Center, NINS, 2-21-1 Osawa, Mitaka, Tokyo 181-8588, Japan}
\affiliation{National Astronomical Observatory of Japan, NINS, 2-21-1 Osawa, Mitaka, Tokyo 181-8588, Japan}

\author[0000-0001-9229-8315]{Konstanze Zwintz}
\affiliation{Institute for Astro- and Particle Physics, University of Innsbruck, Technikerstrasse 25/8, A-6020 Innsbruck, Austria}

\author[0000-0002-7972-0216]{Hiroki Harakawa}
\affiliation{Subaru Telescope, 650 N. Aohoku Place, Hilo, HI 96720, USA}

\author[0000-0002-9294-1793]{Tomoyuki Kudo}
\affiliation{Subaru Telescope, 650 N. Aohoku Place, Hilo, HI 96720, USA}

\author[0000-0002-4677-9182]{Masayuki Kuzuhara}
\affiliation{Astrobiology Center, NINS, 2-21-1 Osawa, Mitaka, Tokyo 181-8588, Japan}
\affiliation{National Astronomical Observatory of Japan, NINS, 2-21-1 Osawa, Mitaka, Tokyo 181-8588, Japan}

\author{Klaus Hodapp}
\affiliation{University of Hawaii, Institute for Astronomy, 640 N. Aohoku Place, Hilo, HI 96720, USA}

\author[0000-0003-1906-4525]{Masato Ishizuka}
\affiliation{Department of Astronomy, Graduate School of Science, The University of Tokyo, 7-3-1 Hongo, Bunkyo-ku, Tokyo 113-0033, Japan}

\author{Shane Jacobson}
\affiliation{University of Hawaii, Institute for Astronomy, 640 N. Aohoku Place, Hilo, HI 96720, USA}

\author{Mihoko Konishi}
\affiliation{Faculty of Science and Technology, Oita University, 700 Dannoharu, Oita 870-1192, Japan}

\author{Takashi Kurokawa}
\affiliation{Astrobiology Center, NINS, 2-21-1 Osawa, Mitaka, Tokyo 181-8588, Japan}
\affiliation{Institute of Engineering, Tokyo University of Agriculture and Technology, 2-24-16, Nakacho, Koganei, Tokyo, 184-8588, Japan}

\author[0000-0001-9326-8134]{Jun Nishikawa}
\affiliation{National Astronomical Observatory of Japan, NINS, 2-21-1 Osawa, Mitaka, Tokyo 181-8588, Japan}
\affiliation{Astrobiology Center, NINS, 2-21-1 Osawa, Mitaka, Tokyo 181-8588, Japan}
\affiliation{Department of Astronomy, School of Science, The Graduate University for Advanced Studies (SOKENDAI), 2-21-1 Osawa, Mitaka, Tokyo 181-8588, Japan}

\author[0000-0002-5051-6027]{Masashi Omiya}
\affiliation{Astrobiology Center, NINS, 2-21-1 Osawa, Mitaka, Tokyo 181-8588, Japan}
\affiliation{National Astronomical Observatory of Japan, NINS, 2-21-1 Osawa, Mitaka, Tokyo 181-8588, Japan}

\author{Takuma Serizawa}
\affiliation{Institute of Engineering, Tokyo University of Agriculture and Technology, 2-24-16, Nakacho, Koganei, Tokyo, 184-8588, Japan}
\affiliation{National Astronomical Observatory of Japan, NINS, 2-21-1 Osawa, Mitaka, Tokyo 181-8588, Japan}

\author{Akitoshi Ueda}
\affiliation{Astrobiology Center, NINS, 2-21-1 Osawa, Mitaka, Tokyo 181-8588, Japan}
\affiliation{National Astronomical Observatory of Japan, NINS, 2-21-1 Osawa, Mitaka, Tokyo 181-8588, Japan}
\affiliation{Department of Astronomy, School of Science, The Graduate University for Advanced Studies (SOKENDAI), 2-21-1 Osawa, Mitaka, Tokyo 181-8588, Japan}

\author[0000-0003-4018-2569]{Sébastien Vievard}
\affiliation{Astrobiology Center, NINS, 2-21-1 Osawa, Mitaka, Tokyo 181-8588, Japan}
\affiliation{Subaru Telescope, 650 N. Aohoku Place, Hilo, HI 96720, USA}

\begin{abstract}
We report the first detection of a hydroxyl radical (OH) emission signature in the planetary atmosphere outside the solar system, in this case, in the day-side of WASP-33b. We analyze high-resolution near-infrared emission spectra of WASP-33b taken using the InfraRed Doppler spectrograph on the 8.2-m Subaru telescope. The telluric and stellar lines are removed using a de-trending algorithm, {\sc SysRem}. The residuals are then cross-correlated with OH and H$_{2}$O planetary spectrum templates produced using several different line-lists. We check and confirm the accuracy of OH line-lists by cross-correlating with the spectrum of GJ 436. As a result, we detect the emission signature of OH at $K_\mathrm{p}$ of 230.9$^{+6.9}_{-7.4}$\,km\,s$^{-1}$ and $v_{\mathrm{sys}}$ of $-$0.3$^{+5.3}_{-5.6}$\,km\,s$^{-1}$ with S/N of 5.4 and significance of 5.5$\sigma$. Additionally, we marginally detect H$_{2}$O emission in the H-band with S/N of 4.0 and significance of 5.2$\sigma$ using the POKAZATEL line-list. However, no significant signal is detected using the HITEMP 2010, which might be due to differences in line positions and strengths, as well as the incompleteness of the line-lists. Nonetheless, this marginal detection is consistent with the prediction that H$_{2}$O is mostly thermally dissociated in the upper atmosphere of the ultra-hot Jupiters. Therefore, along with CO, OH is expected to be one of the most abundant O-bearing molecules in the day-side atmosphere of ultra-hot Jupiters and should be considered when studying their atmosphere.
\end{abstract}


\section{Introduction} \label{sec:intro}
High-resolution spectroscopy is one of the most successful methods to characterize exoplanet atmospheres, especially of hot Jupiters. The resolved planetary lines are disentangled from the telluric and stellar lines due to the planetary orbital motion allowing us to unambiguously detect atomic/molecular signatures in the atmosphere of exoplanets \citep[e.g., ][]{Snellen2010}. By comparing hundreds/thousands of unique absorption/emission lines to model templates through cross-correlation, we can highly boost the signal and constrain the chemical abundances, the planetary rotation, the projected equatorial wind, and even the temperature-pressure (T-P) profile of the atmosphere for emission spectroscopy data.

With the equilibrium temperature (T$_{\mathrm{eq}}$) similar to M-dwarfs, similar prominent atomic/molecular opacity sources (Fe {\sc i}, Fe {\sc ii}, Na {\sc i}, and Ti {\sc ii}, TiO, VO, AlO, FeH, CO, H$_{2}$O, and OH) are expected to be found in the atmosphere of hot Jupiters (T$_{\mathrm{eq}}< 2200$K) and ultra hot Jupiters (T$_{\mathrm{eq}}> 2200$K). Many of these optical opacity sources have been detected \citep[e.g., ][]{Nugroho2017, vonEssen2019, Hoeijmakers2019, Yan2019,  Pino2020, Yan2021}. In the near-infrared, however, the most frequently detected species using high-resolution spectroscopy are CO and H$_{2}$O \citep{Snellen2010, deKok2013, Birkby2013, Lockwood2014, Wang2018, Cabot2019, Webb2020}. The opacities in this wavelength region play an important role as a coolant in the atmosphere and constraining their abundance would allow us to study the climate of unique planetary population and estimate the C/O ratios thus inferring planetary formation history \citep{Oberg2011}. We, therefore, analysed the day-side spectrum of WASP-33b, one of the hottest ultra-hot Jupiters (T$_{\rm{day}}>$3100 K, e.g., \citealt{DeMooij2013}) orbiting a fast-rotating $\delta$-scuti A5-type star \citep{cameron2010}, to search for molecular signatures in the near-infrared.

In this letter, we present the first detection of high-resolution OH emission and the evidence of H$_{2}$O emission in the day-side spectra of WASP-33b. In Section \ref{sec:obs&red}, we describe the observations and data reduction. We then describe our modeling of the planetary emission spectrum in Section \ref{sec:modelspec}, and in Section \ref{sec:crosscor}, we detail our methodology in validating the accuracy of the OH line-lists and searching the signal of OH and H$_{2}$O in the atmosphere of the planet. Finally, in Section \ref{sec:results}, we present and discuss our findings and their implications for the planetary atmosphere.
 
\section{Observations and Data Reductions} \label{sec:obs&red}
We observed WASP-33 on the second half of the night of 30 September 2020 using the InfraRed Doppler instrument \citep[IRD; R\,$\approx$\,70,000; $\lambda $\,$\approx $\,$0.97$-$1.75$\,$\mu$m, ][]{Tamura2012, Kotani2018} on the Subaru 8.2-m telescope (PID: S20B-008, PI: S.K. Nugroho). We continuously observed the target with an exposure time of 300\,s per frame without the laser frequency comb in natural guide star mode. The weather during the observations was not always stable, therefore we were only able to obtain 33 exposures covering the orbital phase of WASP-33b from $\approx$\,0.597 to 0.700 (there were some gaps close to the middle of the observations due to clouds). We converted the Julian Date UTC to Barycentric Julian Date in Barycentric Dynamical Time (BJD$_{\mathrm{TDB}}$) using the online calculator from \citet{Eastman2010} then calculated the orbital phase using the transit epoch taken from \citet{Johnson2015}.

The data were reduced following \citet{Hirano2020} resulting in 70 spectral orders ranging from $\approx$9260-17419 \AA\, with an average signal-to-noise ratio (S/N) of 140. We fitted the continuum of the spectrum with the highest average S/N using {\sc continuum} task in {\sc IRAF}\footnote{The Image Reduction and Analysis Facility ({\sc IRAF}) is distributed by the US National Optical Astronomy Observatories, operated by the Association of Universities for Research in Astronomy, Inc., under a cooperative agreement with the National Science Foundation.} and divided it out from the data. Any possible blaze function variations were then corrected following the procedure in \citet{Nugroho2020MNRAS}. Then, the sky emission lines, bad pixels, and regions were visually identified and masked. Additionally, we also masked any pixels that have a flux less than 10 percent of the continuum. In total, we masked 13.9 percent of the total number of pixels of the data. Finally, the spectra of each spectral order were aligned into 2-dimensional arrays with wavelength along one axis and orbital phase along with the other. We estimated the uncertainty of each pixel by taking the outer product of the standard deviation of each wavelength and exposure bin, then normalized by the standard deviation of the whole array. 

To check if there is any wavelength shift during the observation, the data were cross-correlated with the Doppler-shifted telluric templates produced using the {\sc Cerro Paranal Sky Model} \citep[]{Noll2012, Jones2013} over a velocity range of -50 km s$^{-1}$ to 50 km s$^{-1}$ in 0.01 km s$^{-1}$ steps. We found no significant shift ($<$0.05 km s$^{-1}$) compared to the precision that we need for this analysis, therefore we did not attempt to correct for this.

\begin{figure}
    \centering
    \includegraphics[width=1.0\linewidth]{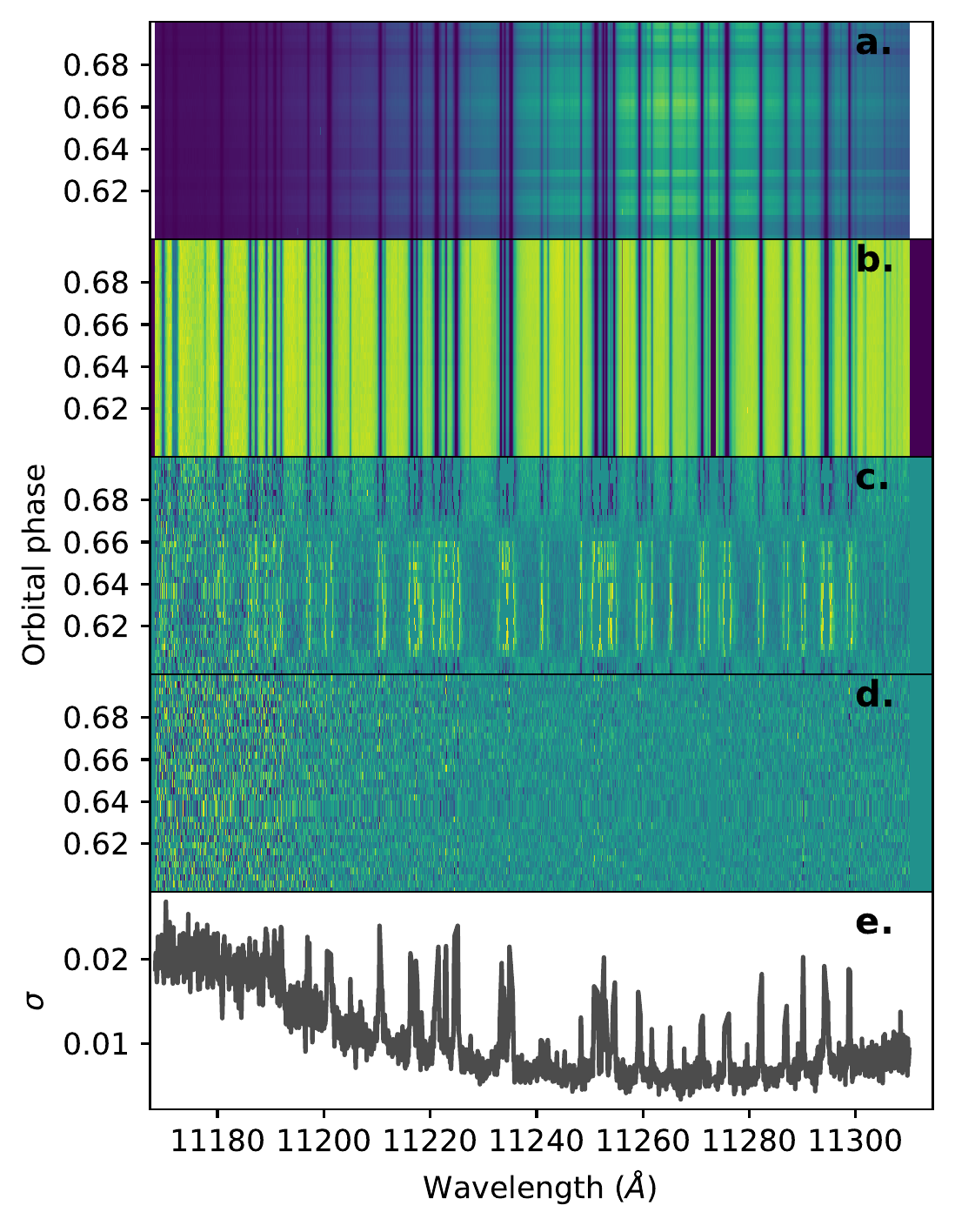}
    \caption{The example of step-by-steps of telluric and stellar line removal for order 26. \textbf{(a.)} The reduced spectra before normalizing and bad-pixels masking. \textbf{(b.)} The normalized reduced spectra after masking bad-pixels and pixels with a value less than 0.1. \textbf{(c.)} The reduced spectra after dividing each wavelength bins by their mean values. \textbf{(d.)} The residual spectra after three iterations of {\sc SysRem}. \textbf{(e.)} The standard deviation ($\sigma$) of each wavelength bin in the residual spectra.
    \label{fig:sysrem}}
\end{figure}

Before searching for any exoplanetary atomic or molecular signal using cross-correlation, we removed the telluric and stellar lines using a de-trending algorithm, {\sc SysRem} \citep{Tamuz2005}, which has been successfully adopted for high-resolution Doppler spectroscopy \citep[e.g., ][]{Birkby2013}. {\sc SysRem} fits the systematic trend in the wavelength bin direction which might be due to variation in airmass, water vapor column level, and others. Following \citet{gibson2020}, we run {\sc SysRem} directly in flux for each spectral order independently. For each iteration, we summed the best-fit {\sc SysRem} model and divided out from the data and the uncertainty array to propagate the error. Finally, any outliers more than five times the standard deviation of the residual were masked. A step-by-step overview of these procedures is shown in Figure \ref{fig:sysrem}. As in the previous analyses using {\sc SysRem} \citep{gibson2020, Merrit2020, Nugroho2020MNRAS, Nugroho2020ApJL, Yan2020}, instead of determining the optimal {\sc SysRem} iteration of each order, we used the same number of iteration for all orders. The results are shown in Section 5.

\section{Planetary Emission Spectrum Templates}
\label{sec:modelspec}

The planetary emission spectrum template was created by assuming 70 atmospheric layers evenly spaced in log pressure from 10$^{2}$ to 10$^{-8}$\,bar of 1D plane-parallel hydro-static atmosphere, and a planetary-mass and radius of 3.266\,M$_{\mathrm{J}}$ and 1.679\,R$_{\mathrm{J}}$, respectively \citep{Kovacs2013}. We adopted a thermally inverted T-P profile used in \citet{Nugroho2020ApJL} which was calculated using the equation in \citet{Guillot2010} assuming the visible mean opacity is twice the infrared mean opacity (0.01 cm$^{2}$g$^{-1}$, e.g., dominated by H$^{-}$ opacity), an internal temperature of 100 K and $T_{\mathrm{eq}}$ of 3100 K (assuming uniform day-side only re-radiation).

We produced 5 planetary spectrum models with single molecular opacity for H$_{2}$O and OH, with three and two different line-lists, respectively. The cross-sections of molecular species were computed using HELIOS-K \citep{Grimm2015} at a resolution of 0.01\,cm$^{-1}$ from 9100 to 17800 \AA assuming a Voigt line profile taking into account natural and thermal broadening only and a line wing cut-off of 100\,cm$^{-1}$. For H$_{2}$O, we used the line-list database of POKAZATEL \citep{Polyansky2018}, HITEMP 2010 \citep{Rothman2010}, and BT2 \citep{Barber2006}; for OH, we used the updated line-list database of HITEMP \citep[the updated HITEMP 2020, ][]{Rothman2010, Brooke2016, Yousefi2018, Noll2020} and MoLLIST \citep{Brooke2016, Yousefi2018, Bernath2020}. For continuum opacity, we included the bound-free and free-free absorption of H$^{-}$ using the equation from \citet{John1988}, and collision-induced absorption (CIA) of H$_{2}$-H$_{2}$ \citep{Abel2011} and H$_{2}$-He \citep{Abel2012}. 

\begin{figure*}
    \centering
    \includegraphics[width=.95\linewidth]{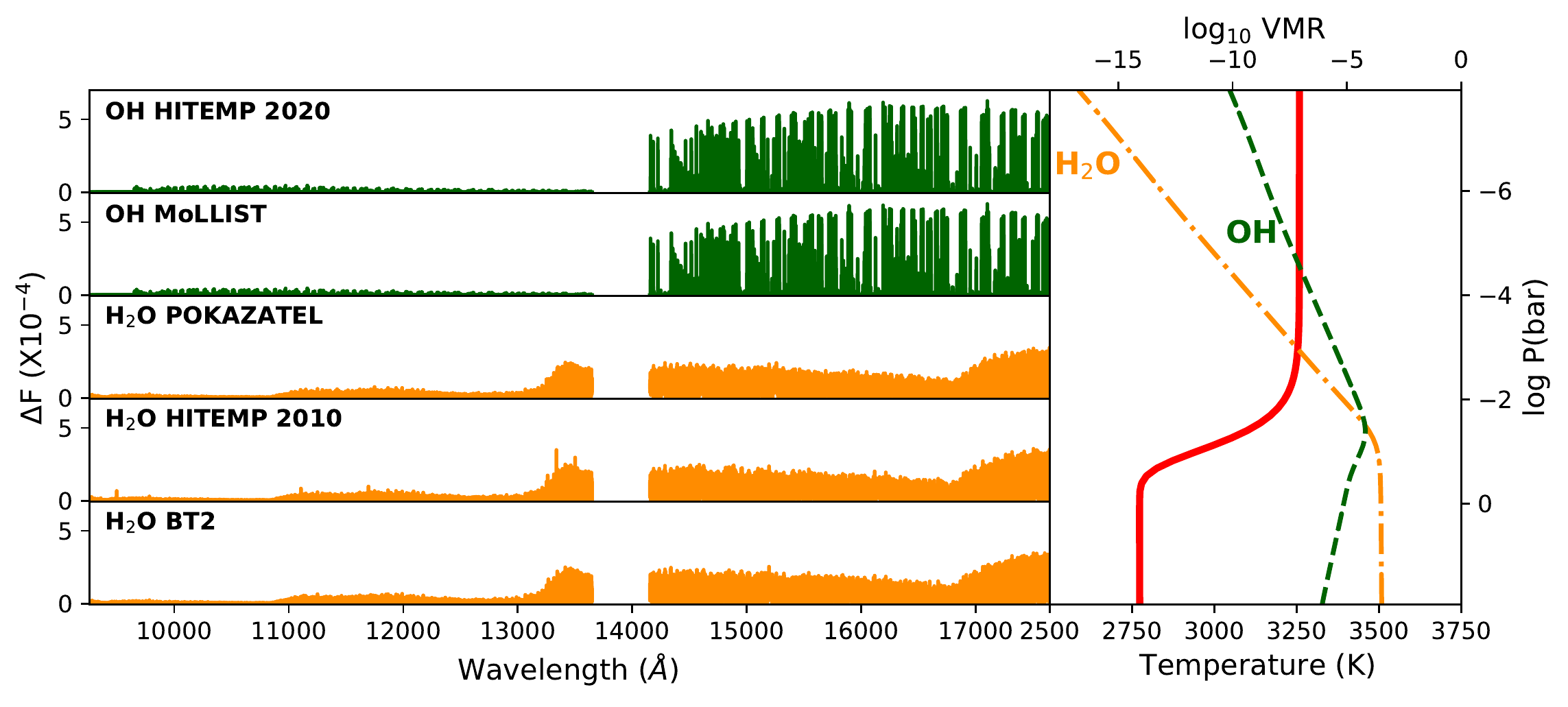}
    \caption{\textit{left panel}: The normalized planetary spectrum models for OH and H$_{2}$O using different line-lists. The scale on the y-axis is the same for all panels, therefore the strength of the emission lines can be compared visually. \textit{right panel}: Temperature-pressure profile of WASP-33b that was adopted in the modeling (red lines). The chemical equilibrium abundances (in volume mixing ratio, VMR) calculated using {\sc FastChem} are indicated by dark green dashed-line for OH and dark orange dotted-line for H$_{2}$O.}
    \label{fig:spectemplate}
\end{figure*}

We used {\sc FastChem} \citep{Stock2018} to estimate the abundances of chemical species, and the mean molecular weight of each atmospheric layer assuming chemical equilibrium and solar C/O. We then produced the emission spectrum by solving the Schwarzchild equation following \citet{Nugroho2017, Nugroho2020ApJL}. We divided the resulting spectra by the flux of the star assuming a black body spectrum, R$_{\star}$ of 1.509\,R$_{\bigodot}$, and $T_{\text{eff}}$ of 7400\,K then convolved with a Gaussian kernel to the spectral resolution of IRD\footnote{Using \textsc{pyasl.instrBroadGaussFast}}. Finally, we subtracted the planetary continuum from each model, which was determined by the continuum opacity of CIA and H$^{-}$, estimated using a minimum filter with a window of 55\,\AA. The final result is the line contrast relative to the stellar continuum profile (see Figure \ref{fig:spectemplate}).

\section{Cross-correlation and Likelihood Mapping}\label{sec:crosscor}
\subsection{Accuracy of the Position of the Lines in the OH Line-lists} \label{subsec:line-list}
In the cross-correlation analysis of high-resolution spectroscopy data, the accuracy of the template, therefore the line-list, has a crucial role in detecting chemical species. Using an incomplete or incorrect line-list might result in biased retrieved parameters \citep{Brogi2019}, or a false-negative detection even when the chemical species exists \citep{Flowers2019}. Following previous analyses \citep[e.g., ][]{Hoeijmakers2015, Nugroho2017}, we investigated the accuracy of the position of the lines in the OH line-lists that we used by cross-correlating the OH planetary spectrum models with GJ 436 spectrum \citep[M2.5V, T$_{\mathrm{eff}}=$ 3416 K, ][]{vonBraun2012} that has a similar temperature as the day-side of WASP-33b. The cross-correlation was done using the Pearson cross-correlation equation over a range of velocity order-by-order. GJ 436's IRD data, taken during the engineering observations, were downloaded from SMOKA \citep{Baba2002}, and reduced in a similar way to our data. Since individual IRD frames are contaminated by telluric lines including air-glow OH emissions, we processed each spectrum and combined multiple frames taken on different epochs to disentangle the stellar lines from the telluric ones based on the procedure described in \citet{Hirano2020}. We should note that the first five spectral orders of the GJ 436 spectrum were still heavily affected by telluric absorption, and thus not included in the cross-correlation analyses below.

The cross-correlation functions of both line-lists are similar in strength and located at the expected radial velocity indicating that they are accurate (see Figure 5 in the Appendix). Most of the lines are located in the $H$-band ($\lambda >$ 1.460 $\mu$m) as was expected for an early-type M-dwarf and our spectrum template (see Figure \ref{fig:spectemplate}). Meanwhile, in the $Y$- and $J$-band, there are no significant or weak correlations which most likely due to weak OH absorption lines in the GJ 436 spectrum. With this result, we concluded that the position of the lines in both line-lists is accurate enough for our purpose.

\subsection[Searching for OH and H2O signatures]{Searching for OH and H$_{2}$O signatures}\label{subsec:OHnH2Osearch}
Even after removing the telluric and stellar lines, the planetary signal is expected to be still buried under the noise. As there are many resolved lines of H$_{2}$O and OH in the IRD wavelength range, we combined them to boost the planetary signal by cross-correlating the residual of each {\sc SysRem} iterations with the Doppler-shifted planetary spectrum templates from $-$500\,km s$^{-1}$ to $+$500\,km\,s$^{-1}$ in 1\,km s$^{-1}$ steps following:
\begin{equation}
    \mathrm{CCF}(v)= \sum_{i} \frac{f_{i}m_{i}(v)}{\sigma_{i}^{2}},
\end{equation}
where $f_{i}$ is the mean-subtracted data, $m_{i}$ is the mean-subtracted spectrum model Doppler-shifted to a radial velocity of $v$, $\sigma_{i}^{2}$ is the variance at $i$th wavelength bin. We performed this for each spectral order and summed them excluding the spectral order that has significant telluric removal residuals (spectral order with central wavelength of 13529.12, 14317.24 and 14457.64 \AA).

We calculated orbital velocity$-$systemic velocity ($K_{\mathrm{p}}-\mathrm{v}_{\mathrm{sys}})$ map by shifting the cross-correlation functions (CCFs) to the planetary rest-frame over a range of $K_{\mathrm{p}}$, from 0 to $+$300\,km s$^{-1}$, and $v_{\mathrm{sys}}$, from $-125$\,km s$^{-1}$ to $+$125\,km s$^{-1}$, both in 0.2\,km\,s$^{-1}$ steps using linear interpolation and summed over time. The radial velocity of the planet at a given orbital phase ($\phi$), RV$_{\mathrm{p}}(\phi)$, assuming the planet has a circular orbit is
\begin{equation}
\mathrm{RV}_{\mathrm{p}} (\phi)= K_{\mathrm{p}} \sin (2\pi\phi) + v_{\mathrm{sys}} + v_{\mathrm{bary}},
\end{equation}where $v_{\mathrm{bary}}$ is the barycentric correction\footnote{using \textsc{pyasl.helcorr}}, and $\phi$ is the orbital phase of the planet. 

From \citet{cameron2010, Nugroho2017, Yan2019, Nugroho2020ApJL}, the planet signal is expected at $K_{\mathrm{p}}$ of $\approx$\,230\,km\,s$^{-1}$ and $v_{\mathrm{sys}}$ of $\approx$\,$-3$\,km\,s$^{-1}$. We computed the S/N by dividing the $K_{\mathrm{p}}-\mathrm{v}_{\mathrm{sys}}$ map by its standard deviation calculated by avoiding the area $\pm$ 25\,km\,s$^{-1}$ from the expected planet signal.

Next, we converted the cross-correlation map to a likelihood map ($\mathcal{L}$) using the $\beta$-optimised likelihood function following \citet[][see also \citealt{Brogi2019}]{gibson2020}:
\begin{equation}\label{eq:lnlikelihood}
    \ln \mathcal{L}= -\frac{N}{2} \ln \left[\frac{1}{N} \left( \sum\frac{f_{i}^{2}}{\sigma_{i}^{2}} + \alpha^{2}\sum \frac{m_{i}^{2}}{\sigma_{i}^{2}}-2\alpha \mathrm{CCF}\right) \right],
\end{equation}
where $\alpha$ is the scale factor of the model and $N$ is the total number of pixels. A 3-dimensional likelihood or posterior data-cube (assuming uniform priors) was then produced with a range of $\alpha$ from 0.01 to 1.50 in 0.01 steps by subtracting the global maximum value from the cube and calculating the exponential, this normalized the likelihood to 1. We then marginalized it by summing the maps over parameters to get the best-fit parameters and uncertainties. We estimated the significance of detection by dividing the median value of the conditional distribution of $\alpha$ at the best-fit value of $K_\mathrm{p}$ and $v_{\mathrm{sys}}$ by its uncertainty.

\section{Results} \label{sec:results}
\subsection{OH emission in the day-side of WASP-33b}\label{sec:OH}
We detected the OH emission signature at S/N of 5.4 and significance of \textbf{5.5$\sigma$} at $K_\mathrm{p}$ of 230.9$^{+6.9}_{-7.4}$\,km\,s$^{-1}$ and $v_{\mathrm{sys}}$ of $-$0.3$^{+5.3}_{-5.6}$\,km\,s$^{-1}$ (see Figure 3b and 4) consistent with previous results although with larger uncertainties due to narrower orbital phase coverage \citep[e.g., ][]{Nugroho2017, Nugroho2020ApJL, Yan2019}. From Figure 3a, the planet signal appears as a bright stripe shown by the white arrows. The strength of the signal varied with time which might be due to the unstable weather during the observation that potentially affects the telluric removal using {\sc SysRem}. The $\alpha$ is constrained to 0.47$\pm0.09$, which means that we have overestimated the strength of the signal. This could be due to the effect of {\sc SysRem} which might have eroded and altered the observed exoplanet signal. Furthermore, the inhomogeneity of the day-side of WASP-33b or the overestimation of the T-P profile and/or the OH abundance in the modeling (photo-dissociation and other possible overlapping opacity sources) could also be the cause. We leave a more detailed analysis to future works.

\begin{figure*}\label{fig:resultccmap}
    \gridline{\fig{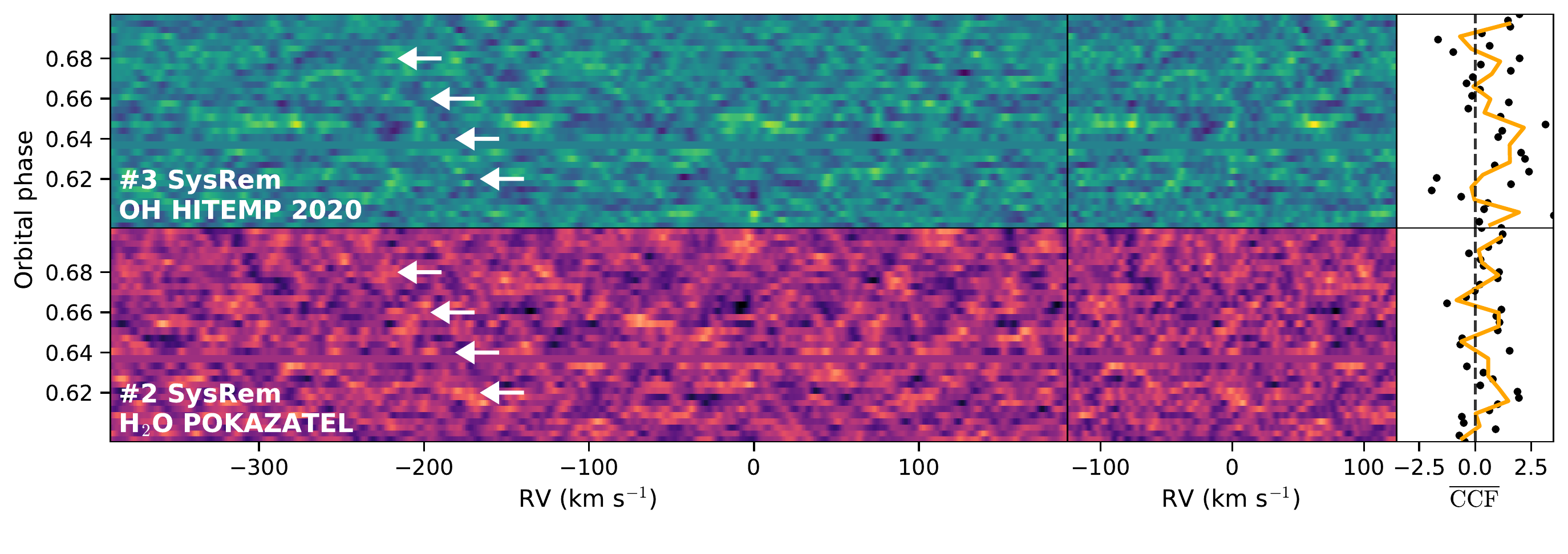}{1\linewidth}{(a)}}
    \gridline{\fig{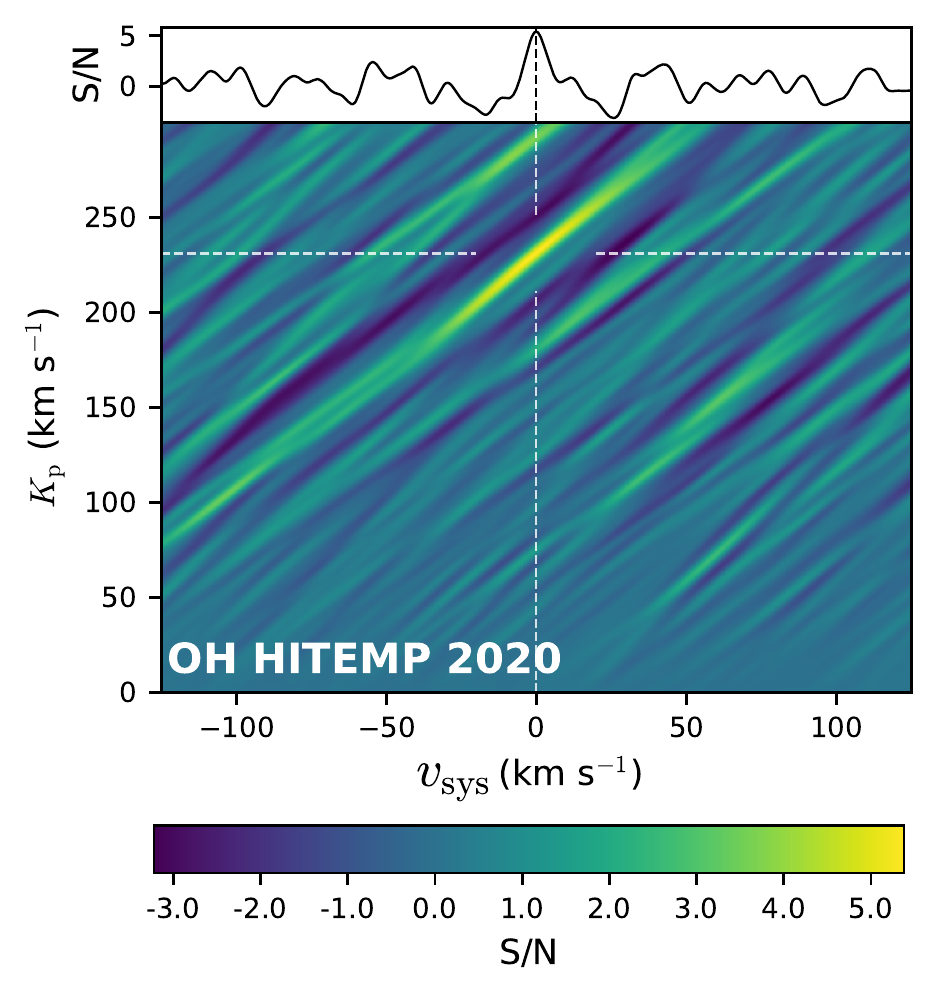}{0.443\linewidth}{(b)}
    \fig{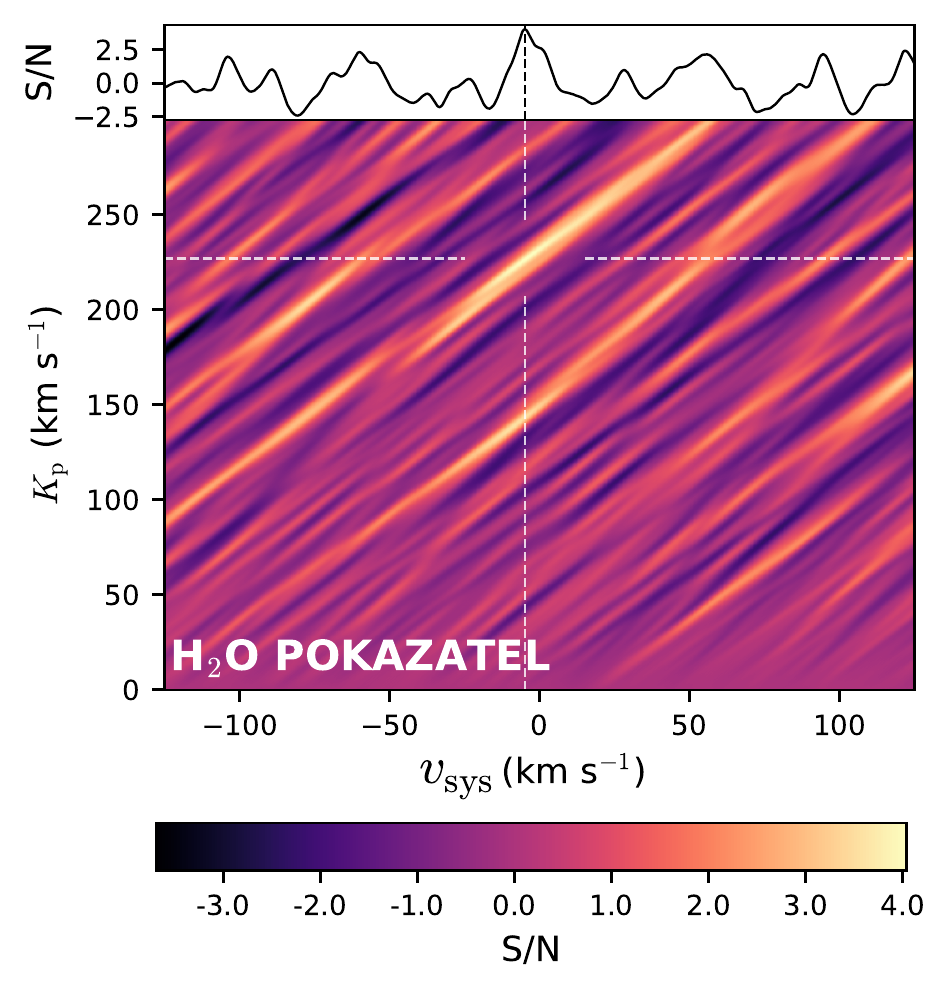}{0.45\linewidth}{(c)}}
    \caption{The cross-correlation results after three {\sc SysRem} iteration for OH HITEMP 2020 (upper panel of \textbf{a}) and after two {\sc SysRem} iteration for H$_{2}$O POKAZATEL (lower panel of \textbf{a}). \textit{left panel}: The cross-correlation map at the telluric rest-frame. The planetary signal appears as a bright diagonal stripe and is shown by the white arrows. \textit{middle panel}: The CCFs at the planetary rest-frame. \textit{right panel}: The mean CCFs of $\pm$ 3\,km s$^{-1}$ from the center of the planet signal are shown by black dots. The orange line shows the binned-CCFs by two exposures. The black dashed line indicates the zero value. The $K_{\mathrm{p}}-\mathrm{v}_{\mathrm{sys}}$ map for the OH HITEMP 2020 \textbf{(b)} and H$_{2}$O POKAZATEL \textbf{(c)}. The white dashed line indicates the maximum signal on the map. The upper panels show the CCFs at the $K_\mathrm{p}$ of 230.9\,km\,s$^{-1}$ and 227.5\,km\,s$^{-1}$ for OH HITEMP 2020 and H$_{2}$O POKAZATEL, respectively. The color-bar shows the S/N of the $K_{\mathrm{p}}-\mathrm{v}_{\mathrm{sys}}$ map.}
\end{figure*}

\begin{figure}\label{fig:likelihood}
    \includegraphics[width=1\linewidth]{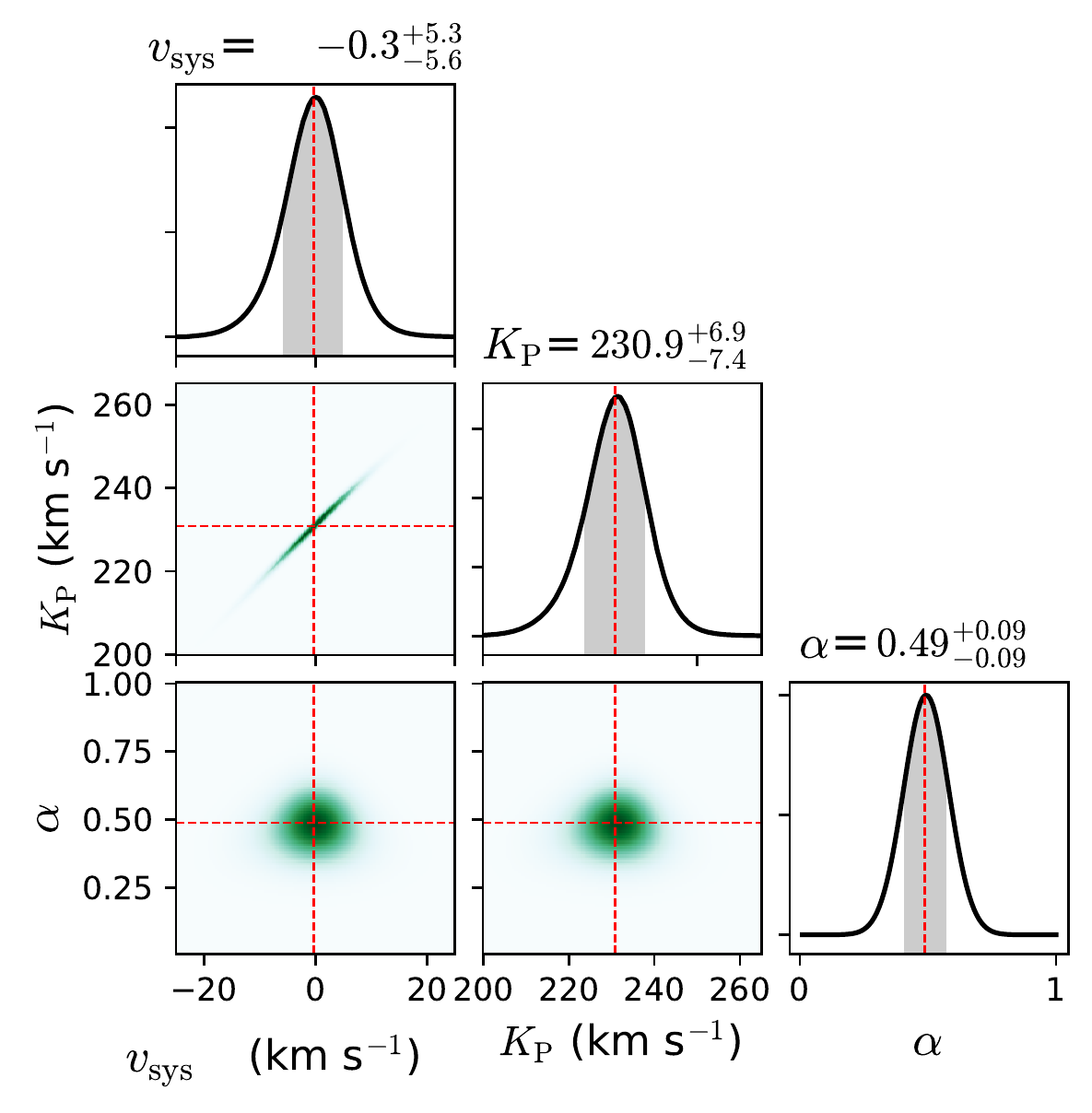}
    \caption{The marginalized likelihood distribution of $K_{\mathrm{p}}$ and $\mathrm{v}_{\mathrm{sys}}$, and conditional likelihood distribution of $\alpha$ at the best-fit $K_{\mathrm{p}}$ and $\mathrm{v}_{\mathrm{sys}}$ for OH HITEMP 2020. The red dashed lines show the median value of the corresponding distribution. The gray shaded area indicates the $\pm$1$\sigma$ limit from the median value.}
\end{figure}

For both line-lists, the detected signals become prominent after two {\sc SysRem} iteration and at their highest S/N after three iterations before getting weaker with more iterations. The cross-correlation maps, the S/N and/or detection significance, and the constraint on alpha for both templates are the same, therefore, we show the result for OH HITEMP 2020 only. The only difference of the result using the two line-lists is in the velocity constraint which differs by $\approx$ 0.1\,km\,s$^{-1}$. This is expected as the two line-lists are based on the same data, although OH HITEMP 2020 line-list was just recently updated based on the observed high-resolution OH telluric lines \citep{Noll2020}. 

Compared to other ultra-hot Jupiters, the atmosphere of WASP-33b's atmosphere is more difficult to characterise with low-resolution spectroscopy/photometry due to the $\delta$-Scuti pulsations of its host star. These pulsations can also affect high-spectral resolution searches for atomic species that occur both in the stellar photosphere and in the planet's atmosphere, such as Fe \,{\sc i} \citep[][Herman et al. in prep.]{Nugroho2020ApJL} where a region of $\sim\pm v_{\mathrm{rot} \star}\,\mathrm{sin}\,i$ has to be excluded from analysis, as the stellar pulsations overlap with the signal from the planet. Since OH is not present in the stellar atmosphere of an A-star, this does not pose a problem for the results presented in this paper. In addition, the lack of signal at $K_\mathrm{p}$ of 0\,km\,s$^{-1}$ or RV of 0\,km\,s$^{-1}$ indicates that there is no contamination from telluric OH emission. Furthermore, as the trail of the signal appeared only at the expected planetary velocity (see Figure 3a and b), we are confident that the detected OH emission signature is originating from the exoplanet.

\subsection[Marginal detection of weak H2O emission?]{Marginal detection of weak H$_{2}$O emission?} \label{sec:H2O}
On the other hand, we only marginally detected H$_{2}$O emission in the H-band at $K_\mathrm{p}$ of 227.5$^{+8.7}_{-8.5}$\,km\,s$^{-1}$ and $v_{\mathrm{sys}}$ of $-$4.3$^{+7.5}_{-6.2}$\,km\,s$^{-1}$ using POKAZATEL line-list at S/N of 4.0 and significance of 5.2$\sigma$. This is consistent with the prediction that most of the molecular feature in the $Y-$ and $J-$band is muted by H$^{-}$ opacity \citep{Arcangeli2018, Parmentier2018}. The S/N of the detected signal may be different to the significance estimated from the conditional likelihood distribution of alpha as they both use different methods to evaluate the noise. For example with the S/N method we have to define a region around the peak to compute the standard deviation, where the conditional likelihood compares to a null signal at the same $K_{\mathrm{p}}$ and $\mathrm{v}_{\mathrm{sys}}$. In this case the lower S/N is likely due to bright spots in the $K_{\mathrm{p}}-\mathrm{v}_{\mathrm{sys}}$ map, and the calculated S/N changes with the arbitrary choice of region used to compute the noise. We argue that the conditional likelihood method is more principled and less arbitrary, but it is nonetheless beneficial to compute the detection significance in multiple ways.

The detected signal is the strongest after two {\sc SysRem} iterations instead of three (see Figure 3c) although it only differs by 0.1 from after two to four {\sc SysRem} iterations. In the OH templates, most of the strongest lines are distributed in the middle of the $H$-band, while for H$_{2}$O, the strong lines are distributed more in both edges of the $H$-band and the redder edge of the $Y$-band. As we performed {\sc SysRem} order-by-order independently, the ''optimum" number of {\sc SysRem} iteration for each order (i.e. optimally removes the telluric lines and leave the planetary signals mostly intact) are potentially different \citep[e.g., ][]{Sanchez2019}. Therefore, the number of {\sc SysRem} iterations that result in the highest S/N of H$_{2}$O signal is potentially to be different than that for the OH signal.

As for HITEMP 2010, the signal around the same location is much weaker (see Figure 6). As we are probing the day-side of the planet, the analysis is sensitive to the T-P profile of the atmosphere. Therefore, following \citet{Nugroho2020ApJL}, we cross-correlated the data with a range of different H$_{2}$O VMR templates assuming uniform abundance with the altitudes to probe different temperatures while minimizing free parameter. However, we found no significant improvement more than our previous analysis. We also tried using BT2 line-list but found no signal (see Figure 6).

\citet{Gandhi2020} also marginally detected H$_{2}$O absorption in the $K$-band from the day-side of HD 179949 using model produced with POKAZATEL line-list (S/N$\approx$ 4.1) and at S/N of 3.2 using HITEMP 2010. For colder hot Jupiter like HD 189733 b, however, both line-lists produced similar detection. These different detections of H$_{2}$O signal could be caused by different line positions and strengths, and the completeness of the line-lists. Thus, even when assuming the same T-P profile and chemical abundances, the results can be different \citep{Brogi2019}. Moreover, at high temperatures, the weak lines can be as abundant as the strong lines thus the completeness of the line-list can be as important as the accuracy of the strong lines. To confirm this, more observational data are needed as we were only able to obtain a marginal detection.

Lastly, we performed an injection test at $K_\mathrm{p}$ of -227.5\,km\,s$^{-1}$ and $v_{\mathrm{sys}}$ of $-$4.3\,km\,s$^{-1}$ with the model produced using the T-P profile and the lower limit of the uniform chemical abundance of H$_{2}$O retrieved by \citet{Haynes2015}, re-ran the {\sc SysRem} then cross-correlated with the template. In contrast to our marginal detection, we were able to recover the injected signal with high significance (S/N$>$ 6). Regardless of whether our detected signal is real or not, this indicates that the retrieved H$_{2}$O abundance and/or the T-P profile in \citet{Haynes2015} might have been overestimated for the upper atmosphere which could be due to the exclusion of the thermal-dissociation effect or OH/H$^{-}$ opacity in the retrieval.

\section{Discussion and Conclusion} \label{sec:conclusion}
While the signature of OH has been detected in the atmosphere of Earth, the Saturn magnetosphere, Venus, and Mars \citep{Meinel1950, Shemansky1993, Piccioni2008, Clancy2013}, this is the first time that its signature has been detected in the atmosphere of an exoplanet. Along with O, OH is one of the most important radical species that drive atmospheric chemistry. For a hot Jupiter like HD 209458b, OH is mainly produced from the photolysis of H$_{2}$O by the stellar UV \citep{Liang2003}. However, for a much hotter planet like WASP-33b, the thermochemical reaction is expected to be the dominant source of OH as the atmosphere is closer to thermochemical equilibrium \citep{Visscher2006}. Our result, which only marginally detected weak emission of H$_{2}$O, indicates that most of the H$_{2}$O in the upper atmosphere is thermally dissociated consistent with the theoretical predictions \citep{Parmentier2018}. Thus, OH is expected to be one of the most dominant O-bearing molecules along with CO and should be considered when analyzing the emission spectrum of ultra-hot Jupiters, as well as search for in other planetary atmosphere. 

Through an injection test, were H$_{2}$O present in the day-side of WASP-33b at the abundance and temperature that were retrieved by \citet{Haynes2015}, we would detect it at high-significance. As low-resolution spectroscopy probes a relatively deeper atmospheric layer than high-resolution spectroscopy, the retrieved parameters might not provide a reliable measurement for the upper atmosphere. Moreover, when there are overlapping unresolved features from multiple species, the retrieved parameters would be incorrect if the model does not consider all of the possible chemical species. Thus, combining low-resolution spectroscopy and high-resolution spectroscopy would be required to get a more accurate and precise characterization of the exoplanet atmosphere \citep{Brogi2019, Gandhi2019}. Finally, this work demonstrates the capability of IRD in characterizing the atmosphere of an exoplanet and its potential to complement the space-borne facilities (e.g., HST, JWST).

\acknowledgements
We are extremely grateful to the anonymous referee for constructive and insightful comments that greatly improved the quality of this letter. This work is based on data collected at Subaru Telescope, which is operated by the National Astronomical Observatory of Japan. Our data reductions benefited from (PyRAF and) PyFITS that are the products of the Space Telescope Science Institute, which is operated by AURA for NASA. The M-dwarfs spectra are based on data collected at Subaru Telescope and obtained from the SMOKA, which is operated by the Astronomy Data Center, National Astronomical Observatory of Japan. H.K. is supported by a Grant-in-Aid from JSPS (Japan Society for the Promotion of Science), Nos. JP18H04577, JP18H01247, and JP20H00170. This work was also supported by the JSPS Core-to-Core Program "Planet$^{2}$" and SATELLITE Research from Astrobiology Center (AB022006). N. P. G. gratefully acknowledges support from Science Foundation Ireland and the Royal Society in the form of a University Research Fellowship. T.H. acknowledges support from JSPS KAKENHI grant number 19K14783. Y.K. is supported by Special Postdoctoral Researcher Program at RIKEN. C.A.W. would like to acknowledge support from UK Science Technology and Facility Council grant ST/P000312/1.  M.T. would like to acknowledge support from MEXT/JSPS KAKENHI grant Nos. 18H05442, 15H02063, and 22000005. M.I. acknowledges support from JSPS KAKENHI grant numbers 19J11805. JLB acknowledges funding from the European Research Council (ERC) under the European Union’s Horizon 2020 research and innovation program under grant agreement No 805445. MB acknowledges support from the UK Science and Technology Facilities Council (STFC) research grant ST/S000631/1. We are also grateful to the developers of the {\sc Numpy}, {\sc Scipy}, {\sc Matplotlib}, {\sc Jupyter Notebook}, and {\sc Astropy} packages, which were used extensively in this work \citep{2020SciPy-NMeth, Hunter:2007, Kluyver:2016aa, astropy:2013, astropy:2018}.

\appendix
\section{Additional Figures}
\begin{figure*}\label{fig:line-listcc}
    \centering
    \includegraphics[width=1.0\linewidth]{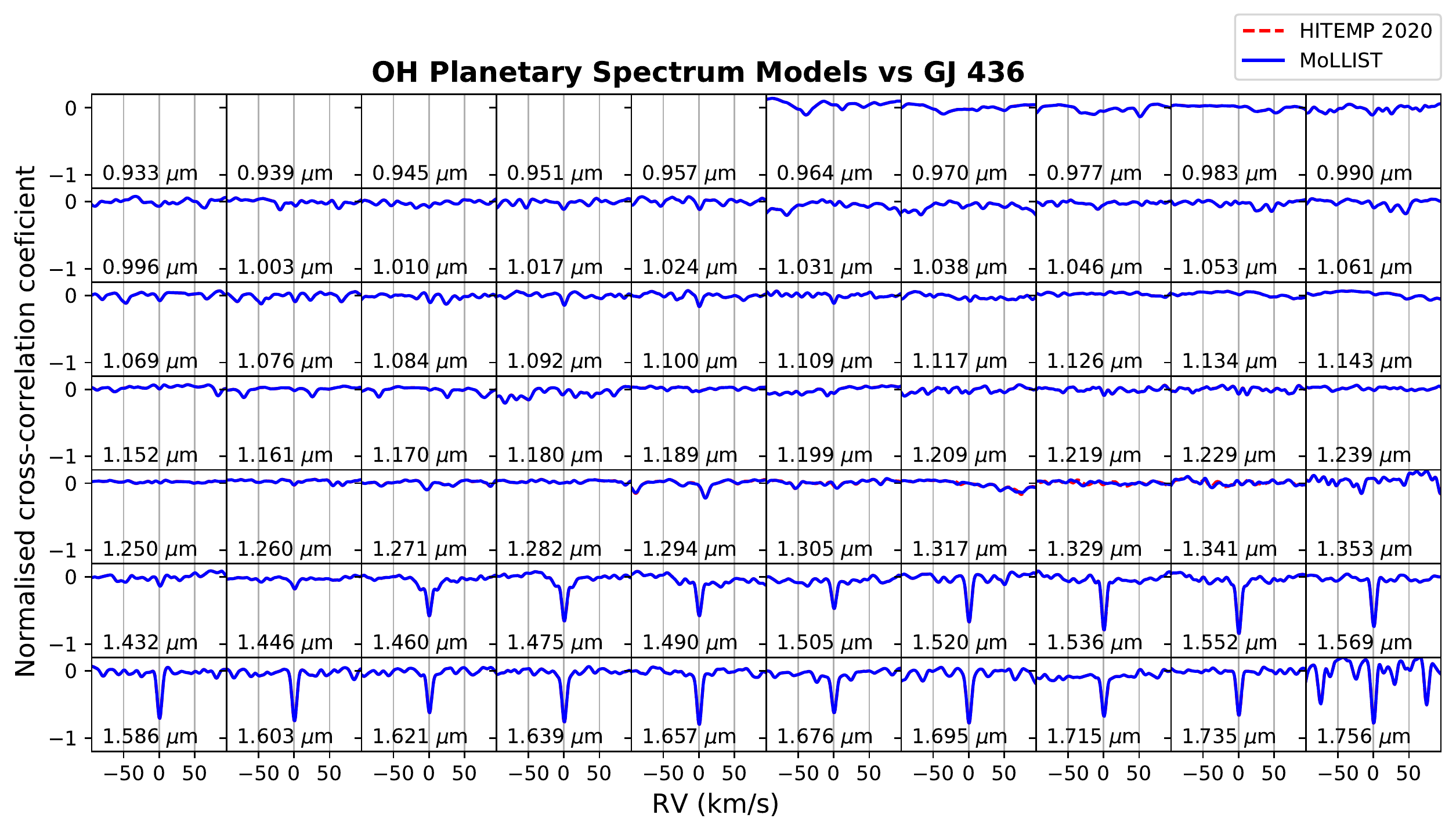}
    \caption{The CCFs between GJ 436 spectrum and OH planet spectrum models using HITEMP 2020 (red dashed-line) and MoLLIST (blue line) for each spectral order (each panel labeled by the median wavelength value of each spectral orders). The CCFs of HITEMP 2020 overlap with the CCFs of MoLLIST indicating that they have the similar accuracy.}
\end{figure*}

\begin{figure*}\label{fig:h2o_other_ccmap}
    \gridline{\fig{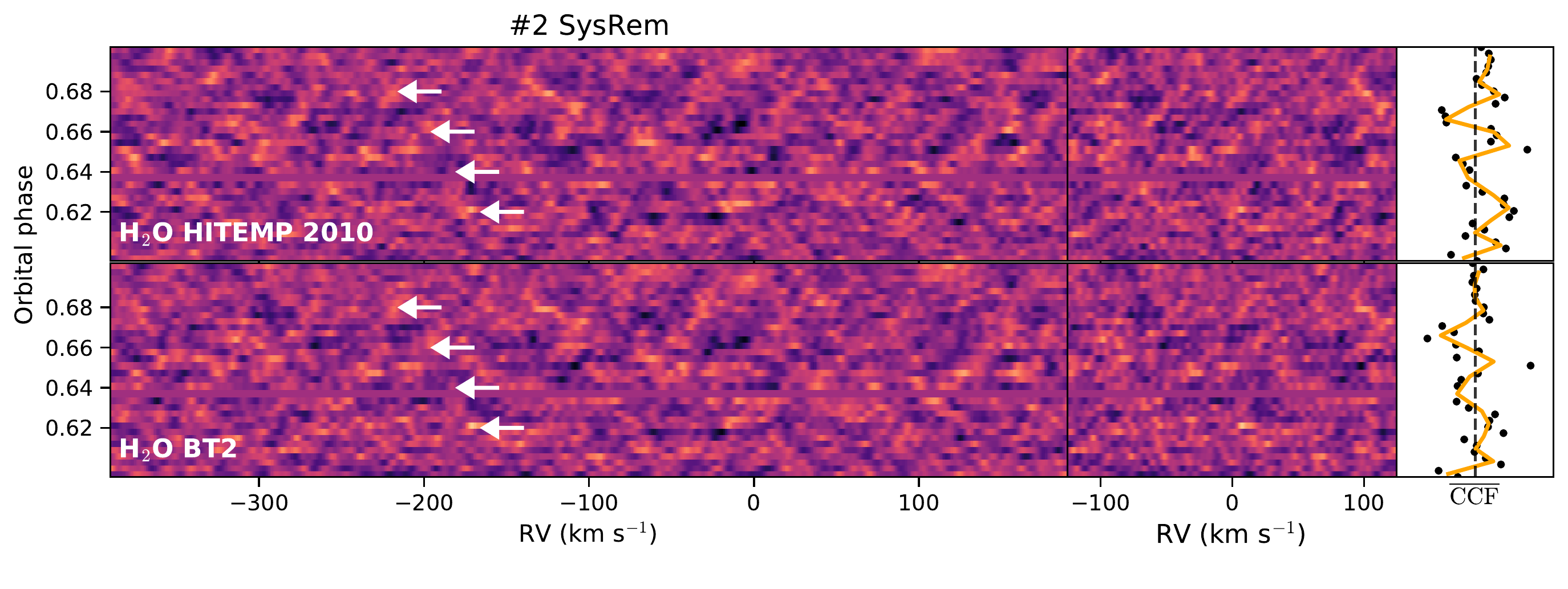}{1\linewidth}{(a)}}
    \gridline{\fig{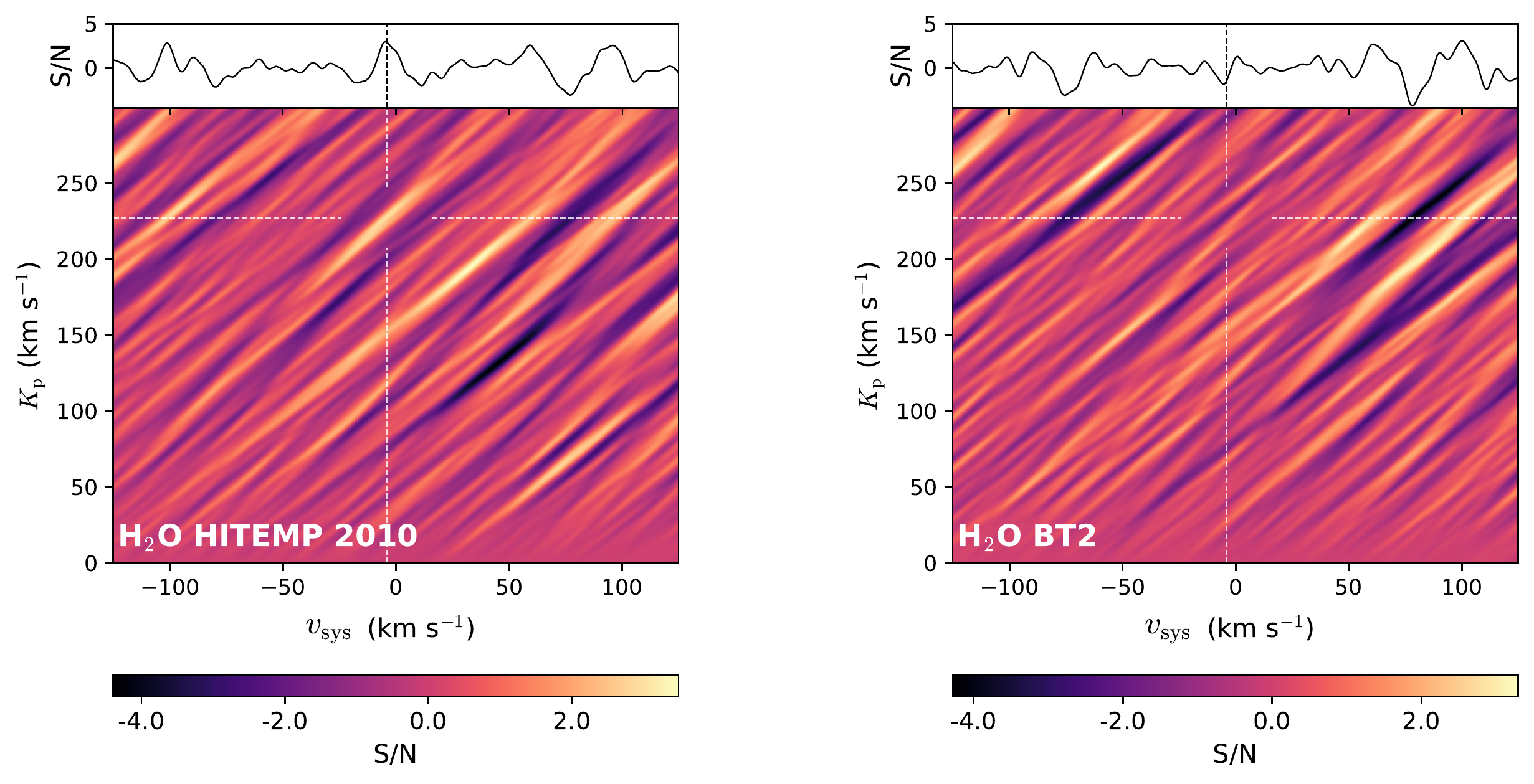}{1\linewidth}{(b)}}
    \caption{Similar to Figure 3 but for HITEMP 2010 and BT2. The white dashed line is fixed at the $K_\mathrm{p}$ of 227.5\,km\,s$^{-1}$ and $v_{\mathrm{sys}}$ of $-$4.3\,km\,s$^{-1}$. The upper panels show the CCFs at the same $K_\mathrm{p}$.}
\end{figure*}

\bibliography{OHW33b-nugroho}{}
\bibliographystyle{aasjournal}



\end{document}